\documentclass{jpsj2}
\title{Fluctuation of the Top Location and Avalanches in the Formation
Process of a Sandpile}

\author{Chiyori \textsc{Urabe}\thanks{E-mail adress:
chiyori@scphys.kyoto-u.ac.jp}}

\inst{Graduate School of Science, Kyoto University, Kyoto 606-8501}

\abst{

We investigate the formation processes of a sandpile using numerical
 simulation.
We find a new relation between the fluctuation of the motion of the top
and the surface state of a sandpile.
The top moves frequently as particles are fed one by one every time
 interval $T$.
The time series of the top location has the power spectrum which obeys a
power law, $S(f)\sim f^{\alpha}$, and its exponent $\alpha$ depends on $T$
and the system size $w$.
The surface state is characterized by two time scales;
the lifetime of an avalanche, $T_{a}$, and the time required to cause an
 avalanche, $T_{s}$.
The surface state is fluid-like when $T_{a}\sim T_{s}$, and it is
solid-like when $T_{a}\ll T_{s}$.
Our numerical results show that $\alpha$ is a function of $T_{s}/T_{a}$.

}

\kword{sandpile, fluctuation, the top location, power spectrum,
power law, avalanche, fluid state, solid state, granular system,
numerical simulation}

\begin{document}
\maketitle

\section{Introduction} %% No sections necessary for express letters, letters
                       %% and short notes

It is known that the state of a granular system changes between solid-like
state and fluid-like state
\cite{
NeddermanB1992,
JaegerNagelBehringerRMP68(1996), 
KadanoffRMP71(1999),
DuranB2000,
PoschelLuding2001,
PoschelBrilliantov2003
}.
A sandpile in the formation process is a typical system in which the
both states appear.
Its state is solid-like when the feed rate of particles to
the sandpile is sufficiently small, and the stress in the sandpile
localizes in certain particles
\cite{
WittmerClaudinCatesBouchaudN382(1996),
VanelHowellClarkBehringerClementPRE60(1999),
GengHowellLonghiBehringerPRL87(2001),
GengLonghiBehringerHowellPRE64(2001)
}.
Contrastingly, the surface of the sandpile is in fluid-like state when the
feed rate is large because avalanches occur frequently, and it is
reported that the magnitude distribution of avalanches depends on the
grain shape and the system size
\cite{
Nature379_49,
AltshulerRamos_etalPRL86(2001),
YoshiokaEarthPlanetsSpace55(2003)
}.
In particular, the surface state varies locally and with time, and it
is controlled by the feed rate.
Elucidation of the transition of the state is one of interesting
problems in granular systems.
%--------------------

In the previous paper\cite{UrabeJPSJ74(2005)}, we studied numerically the
formation process of a two-dimensional sandpile.
We found that the power spectrum of the time series of the top location,
$S(f)$, obeys generically a power law, $S(f)\sim f^{\alpha}$, and that
the exponent $\alpha$ depends on the feed rate.
We defined the left(right) mode as the state which avalanches occur
mainly on the left(right) slope of the sandpile.
In the case where the feed rate is large, if we introduce a two-valued
function $K$ which takes $-1$ when the left mode appears and $1$ when the
right mode does, the power spectrum of the time series of $K$ obeys a
power law, and its exponent is equal to $\alpha$.
%--------------------

In this paper, we investigate the fluctuation of the top location and
avalanches in two-dimensional and three-dimensional sandpiles in more detail.
Avalanches occur on the surface of the sandpile, and the surface state is
characterized by two time scales;
the lifetime of an avalanche, $T_{a}$, and the time required to cause an
avalanche, $T_{s}$.
We find that the surface state which changes between fluid-like state and
solid-like state and $\alpha$ depend on
$T_{s}/T_{a}$ for a two-dimensional sandpile.
We redefine the left(right) mode as the state that almost avalanches in a
time interval occur on the left(right) slope and observe the
continuation of the left or right mode for sufficiently long time.
We find that the reason of the continuation is that the memory of the
mode is stored in the shape of the sandpile.
In addition, also for a three-dimensional sandpile, the power spectrum of the
time series of the top location obeys a power law with the exponent which
depends on the feed rate.
%--------------------

This paper is organized as follows.  
In the next section, we describe the simulation method and the setup
of the system.
In Sec.3, for the two-dimensional sandpile, it is shown that the power
spectrum of the top location obeys a power law with the exponent
$\alpha$ which depends on $T$ and $w$.
In Sec.4, we consider avalanches in the two-dimensional sandpile.
In Sec.5, the relations among $T_{a}$,$T_{s}$ and $\alpha$ are discussed.
In Sec.6, we present the results for the three-dimensional sandpile.
Sec.7 is devoted to discussion and summary.
%---------------

\section{Discrete Element Method}
We numerically simulate the motion of particles using the discrete
element method (DEM)~\cite{CundallStrackG29(1979)}.
Particles are circular in a two-dimensional system or
spherical in a three-dimensional system with the radii uniformly
distributed in the range $[0.8d,d]$.
The force of gravity acts on every particle, and elastic force, viscous
force and coulomb friction affect each pair of particles in contact.
Let $m_{i}$, $I_{i}$ and $r_{i}$ denote the weight, the moment of inertia and
the radius of the $i$th particle, respectively.
The center of mass, $\mathbf{x}_{i}$, and the angular velocity
$\omega_{i}$ of the $i$th particle obey the following equations of
motion.
\begin{eqnarray} 
m_{i}\ddot{\mathbf{x}}_{i}&=&\sum_{j}\Theta(X_{ij})(F_{n}^{ij}\mathbf{n}_{ij}+\mathbf{F}_{t}^{ij})+m_{i}\mathbf{g},
\label{dem1}\\
I_{i}\dot{\mathbf{\omega}}_{i}&=&r_{i}\sum_{j}\Theta(X_{ij})\mathbf{n}_{ij}\times \mathbf{F}_{t}^{ij}, 
\label{dem2}
\end{eqnarray}
where $\Theta$ is the Heaviside function, and $\mathbf{n}_{ij}$ and $X_{ij}$
are defined as 
\[
\mathbf{n}_{ij}=(\mathbf{x}_{j}-\mathbf{x}_{i})/\mid\mathbf{x}_{j}-\mathbf{x}_{i}
\mid,
\]
and
\[
X_{ij}=r_{i}+r_{j}-\mid \mathbf{x}_{i}-\mathbf{x}_{j}\mid ,
\]
respectively.
The normal contact force $F_{n}^{ij}\mathbf{n}_{ij}$ and the tangential
contact force $\mathbf{F}_{t}^{ij}$ are calculated as follows. 
We define $F_{n}^{ij}$  as
\begin{eqnarray}
F_{n}^{ij}&=&\tilde{F}_{n}^{ij}\Theta(-\tilde{F}_{n}^{ij}),\label{dem3}
\end{eqnarray}
where 
\begin{eqnarray}
\tilde{F}_{n}^{ij}&=&-k_{n}X_{ij}-\eta_{n}\mathbf{n}_{ij}\cdot
 \left(\dot{\mathbf{x}}_{i}-\dot{\mathbf{x}}_{j}\right),\label{dem4}
\end{eqnarray}
commonly in the two and three dimensional systems.
The function $\Theta (-\tilde{F}_{n^{ij}})$ means that particles are
cohesionless.
Parameters $k_{n}$ and $\eta_{n}$ represent the spring constant and the viscous
coefficient in the normal direction.

%---------------

We employ different definition of $\mathbf{F}_{t}^{ij}$ in the two and
three dimensional systems.
In the two-dimensional system, $\mathbf{F}_{t}^{ij}$ is defined as in the
previous paper~\cite{UrabeJPSJ74(2005)}, 
\begin{equation}
\mathbf{F}_{t}^{ij}=k_{t}u_{t}^{ij}\mathbf{t}_{ij},
\end{equation}
where  $\mathbf{t}_{ij}$ is the tangential vector, and $k_{t}$ is the
spring constant in the tangential direction.
Displacement $u_{t}^{ij}$ is given by the integration of the
following equation under the condition that the $i$th and $j$th
particles are in contact, that is when $|\mathbf{x}_{j}-\mathbf{x}_{i}|
\leq r_{j}+r_{i}$.
\begin{equation}
\dot{u}_{t}^{ij}=-\Bigl( (\dot{\mathbf{x}}_{i}-\dot{\mathbf{x}}_{j})\cdot
\mathbf{t}_{ij}+r_{i}\omega_{i}+r_{j}\omega_{j}\Bigr)
\Theta\bigl(\mu|F_{n}^{ij}|-|F_{t}^{ij}| \bigr),
\end{equation}
where $\mu$ is the friction coefficient, and $u_{t}^{ij}$ is zero when
$|\mathbf{x}_{j}-\mathbf{x}_{i}| > r_{j}+r_{i}$.
In the three-dimensional system, $\mathbf{F}_{t}^{ij}$ is
defined as follows.
\begin{eqnarray}
\mathbf{F}_{t}^{ij}=
\left\{
\begin{array}{cc}
\tilde{\mathbf{F}}_{t}^{ij} & \mbox{if} \mid
\tilde{\mathbf{F}}_{t}^{ij}\mid<\mu \mid F_{n}^{ij}\mid, \\
\mu F_{n}^{ij} \mathbf{e}_{t}^{ij} & \mbox{otherwise},
\end{array}
\right. 
\end{eqnarray}
where
\[
\tilde{\mathbf{F}}_{t}^{ij}=-k_{t}\mathbf{\Psi}-\eta_{t}\left(
\mathbf{n}_{ij}\times \left(\dot{\mathbf{x}}_{j}-\dot{\mathbf{x}}_{i}\right)
+r_{i}\mathbf{\omega}_{i}+r_{j}\mathbf{\omega}_{j}\right)
\times\mathbf{n}_{ij},
\]
\[
\mathbf{\Psi}=\sum_{l=1}^{2}\mathbf{t}_{l}
\int_{t_{0}}^{t}dt ' \tilde{\mathbf{\Psi}}(t ' )\cdot \mathbf{t}_{l}(t ' ),
\]
\[
\tilde{\mathbf{\Psi}}(t ' )=
\left(r_{i}\mathbf{\omega}_{i}+r_{j}\mathbf{\omega}_{j}\right)\times\mathbf{n}_{ij}(t ' )
+\dot{\mathbf{x}}_{j}(t ' )-\dot{\mathbf{x}}_{i}(t ' ),
\]
\[
\mathbf{e}_{t}^{ij}=\frac{\mathbf{F}_{t}^{ij}}{\mid\mathbf{F}_{t}^{ij}\mid}.
\]
Time $t_{0}$ is the time when the $i$th and $j$th particles begin to contact. 
Parameter $\eta_{t}$ is the viscous coefficient in the tangential direction.
The tangential vectors, $\mathbf{t}_{1}$ and $\mathbf{t}_{2}$, are unit
vectors perpendicular to $\mathbf{n}_{ij}$. 

%----------------------

There are two physical differences for the tangential forces in the
two and three dimensional systems.
One is that viscous term is absent in the two-dimensional
system, while it is present in the three-dimensional system.
Although the latter describes more general cases, we emphasize
consistency with the previous work~\cite{UrabeJPSJ74(2005)}.
The other is that the tangential shear has the maximum value in the
two-dimensional system, while there is no limit on the shear in the
three-dimensional system. 
In the two-dimensional system, we assume that the shear beyond the
limit does not affect because particles slip. 
Although these differences influence significantly dense systems with
strong shear acting continuously, we infer that the influence is
noncritical for our sandpile systems because the contact time is not
very long.

%----------------------

Parameter values used in our simulation are listed in Table \ref{para},
and the physical quantities are rescaled to be dimensionless where $m$
represents the weight of a particle with radius $d$.
The restitution constants in two and three dimensional systems are
respectively about $0.3$ and $0.2$ with the values in Table \ref{para}.

%----------------------

\begin{figure}
\begin{center}
\includegraphics[width=7.5cm]{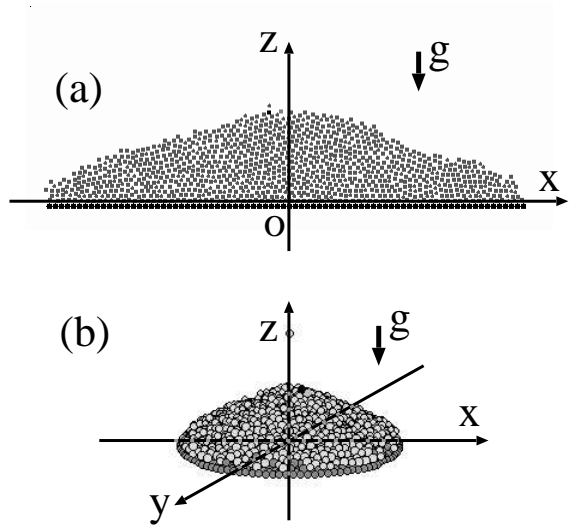}
\caption{\label{sp}The two-dimensional sandpile for $w=80d$ (a) and (b) the
 three-dimensional sandpile for $w=30d$ (b).}
\end{center}
\end{figure}

We make a sandpile on a table which has the origin of coordinates at
the center.
The table in the two-dimensional system is illustrated in Fig. \ref{sp}
(a).
It consists of an alignment of particles with diameter $d$ on the
$x$-axis, and its length is $w$.
In the three-dimensional system, the table is a flat circular plate with
diameter $w$ on the $xy$ plane as shown in Fig.\ref{sp}(b).
It has fixed particles with diameter $0.8d$ on its fringe.
The contact force between particles and the plate is calculated in the
same manner as that between two particles. 

%----------------------

We carry out simulations using a initial sandpile which is large
to cover the table.
The size of a sandpile is kept virtually constant because particles are
eliminated if they fall from the table.
For a three-dimensional initial sandpile, after we feed sufficiently many
particles and make an initial sandpile, we fix the particles remained in
the sandpile for a long time to reduce calculation cost.
We use Adams-Bashforth method to calculate the time evolution of particles,
and the time step is $\delta t = 1.0\times 10^{-3}$.  

%----------------------

\begin{table}[t]
\caption{\label{para} Variables and Parameters}
\begin{center}
 \begin{tabular}{ccc} \hline
& 2D & 3D \\  \hline
$m_{i}$ & $m\left(2r_{i}/d\right)^{2}$ & $m\left(2r_{i}/d\right)^{3}$\\
$I_{i}$ & $ m_{i}r_{i}^{2}/2$ & $2m_{i}r_{i}^{2}/5 $ \\
  \hline
$k_{n}[mg/d]$ & $1.0\times 10^{4}$ & $1.0\times 10^{4}$  \\
$\eta_{n}[m\sqrt{d/g}]$ & $1.0\times 10^{2}$ & $1.4\times10^{2}$ \\
$k_{t}[mg/d]$ & $2.0\times 10^{3}$ & $2.5\times 10^{3}$ \\
$\eta_{t}[m\sqrt{d/g}]$ & $0$ & $7.2\times 10$ \\
$\mu$ & $0.5$ & $0.2$ \\ \hline
$w$ & $20d,40d,80d,160d$ & $30d$ \\ \hline
\end{tabular}
\end{center}
\end{table}

%----------------------

We feed particles to the sandpile as follows.
Particles are dropped one by one from above the origin every time
interval $T$ whether avalanches occur or not.
The height from which particles are dropped, $H$, is measured from the
surface in the two-dimensional sandpile, and $H$ is fixed to keep the
collision impact given by dropped particle constant.
Thus, the height from the table changes with time.
In the three-dimensional sandpile, we fix the height from the table,
$H'$,  for facilitation of experiments with the same setting. 

%----------------------

%----------------------

\section{Fluctuation of the Top Location in the Two-Dimensional Sandpile}

The top location is defined as the center of mass of the highest
particle in contact with others.
This quantity indicates the shape of a sandpile, and the top is moved by
avalanches.

%----------------------

we measure the horizontal position of the top, $x_{top}$, in the
two-dimensional sandpile and calculate the power spectrum of the time series
of $x_{top}$ for given parameter set $(w,H,T)$ to characterize the motion of
the top.
For each parameter set, the power spectrum $S(f)$ is obtained through
the sample average, and the average is over the power spectra of more
than ten time series with the length $10^{4}\sqrt{d/g}$.  
In the previous paper~\cite{UrabeJPSJ74(2005)}, only for $w=80d$, we found
that $S(f)$ obeys a power law, $S(f)\sim f^{\alpha}$, and that the exponent
$\alpha$ increases as $T$ decreases, while $\alpha$ is independent of $H$.

%----------------------

\begin{figure}
\begin{center}
\includegraphics[width=7.5cm]{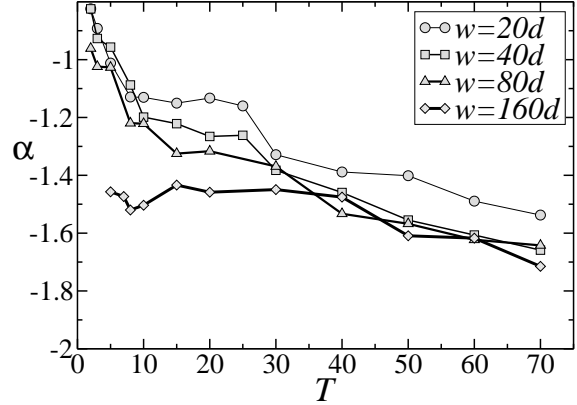}
\caption{\label{a2D} Dependence of $\alpha$ on $T$ and $w$ for $H=20d$}
\end{center}
\end{figure}

In this paper, we look into the dependence of $\alpha$ on $w$.
Exponent $\alpha$ is calculated by applying the least-squares
method for the double logarithmic plot of $S(f)$ in the frequency range
$0.0005<f<0.01$.
Fig.\ref{a2D} shows that $\alpha$ tends to decrease with $w$ and increases
drastically as $T$ decreases in the range $T<10\sqrt{d/g}$ not only when
$w=80d$ but also in cases where $w=20d$ and $40d$.
For $w=160d$, however, the range in which $\alpha$ changes with $T$ is small.
In Sec.5, we consider again the dependence of $\alpha$ on
$w$ through the relation between $\alpha$ and avalanches.

%----------------------

%----------------------

\section{Avalanches in the Two-Dimensional Sandpile}

\subsection{Avalanches for small $T$}

%----------------------

\begin{figure}
\begin{center}
\includegraphics[width=7.5cm]{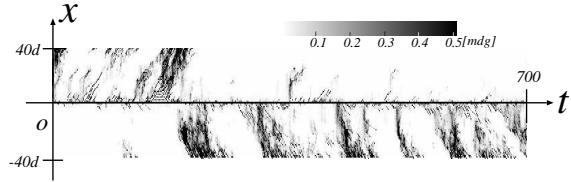}
\caption{\label{ts} The time-space plot of kinetic energy for
 $(w,H,T)=(80d,20d,2\sqrt{d/g})$.}
\end{center}
\end{figure}

%----------------------

To observe avalanches, we show the time-space plot of kinetic energy for small
$T$ in Fig.\ref{ts}.
The gray scale represents the value of kinetic energy of particles at position
$x$.
The kinetic energy is high at $x=0$ because particles land there.
Avalanches continue for a long time on the left slope which is the lower half
in Fig.\ref{ts}, and the duration time is sufficiently large in
comparison with $T$.
%-----------------

In the previous paper \cite{UrabeJPSJ74(2005)}, to measure avalanches on
each slope, we calculated kinetic energy of particles in the left or
right half of the sandpile.
The kinetic energy in each part defined as the range $x>d$ or $x<-d$ is denoted
by $k_{l}$ or $k_{r}$.
The magnitude relation between $k_{l}$ and $k_{r}$ changes with time.
We defined $K(t)$ as
\begin{eqnarray}
K(t)=\left\{
\begin{array}{cc}
1, \quad & \mbox{if $k_{l}(t)<k_r(t)$,} \\
-1, \quad & \mbox{otherwise,}
\end{array}
\right.\label{K}
\end{eqnarray}
and we called the state that $K=1$($K=-1$) the right(left) mode.
For small $T$, the exponents of the power spectra of $x_{top}$ and $K$
are approximately equal in a low frequency range.

%------------------

\subsection{Redefinition of the mode}

%------------------

To investigate the mode in more detail, we redefine the mode
which depend on the state of avalanches for a long time scale.
Because the current definition determined by instantaneous
magnitude of avalanche, it is not able to operate when avalanches occur
intermittently.

%----------------------
\begin{figure}
\begin{center}
\includegraphics[width=7cm]{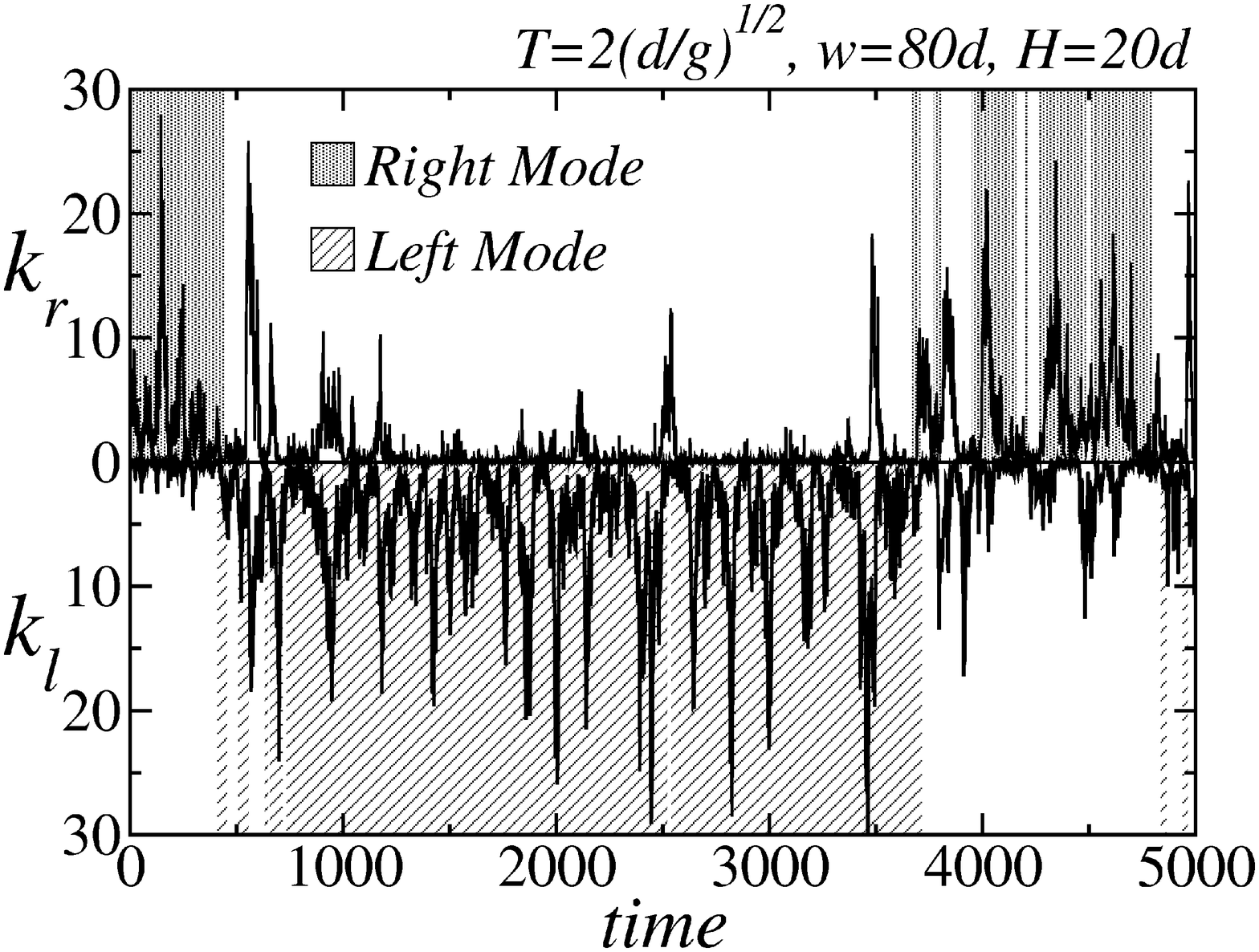}
\includegraphics[width=7cm]{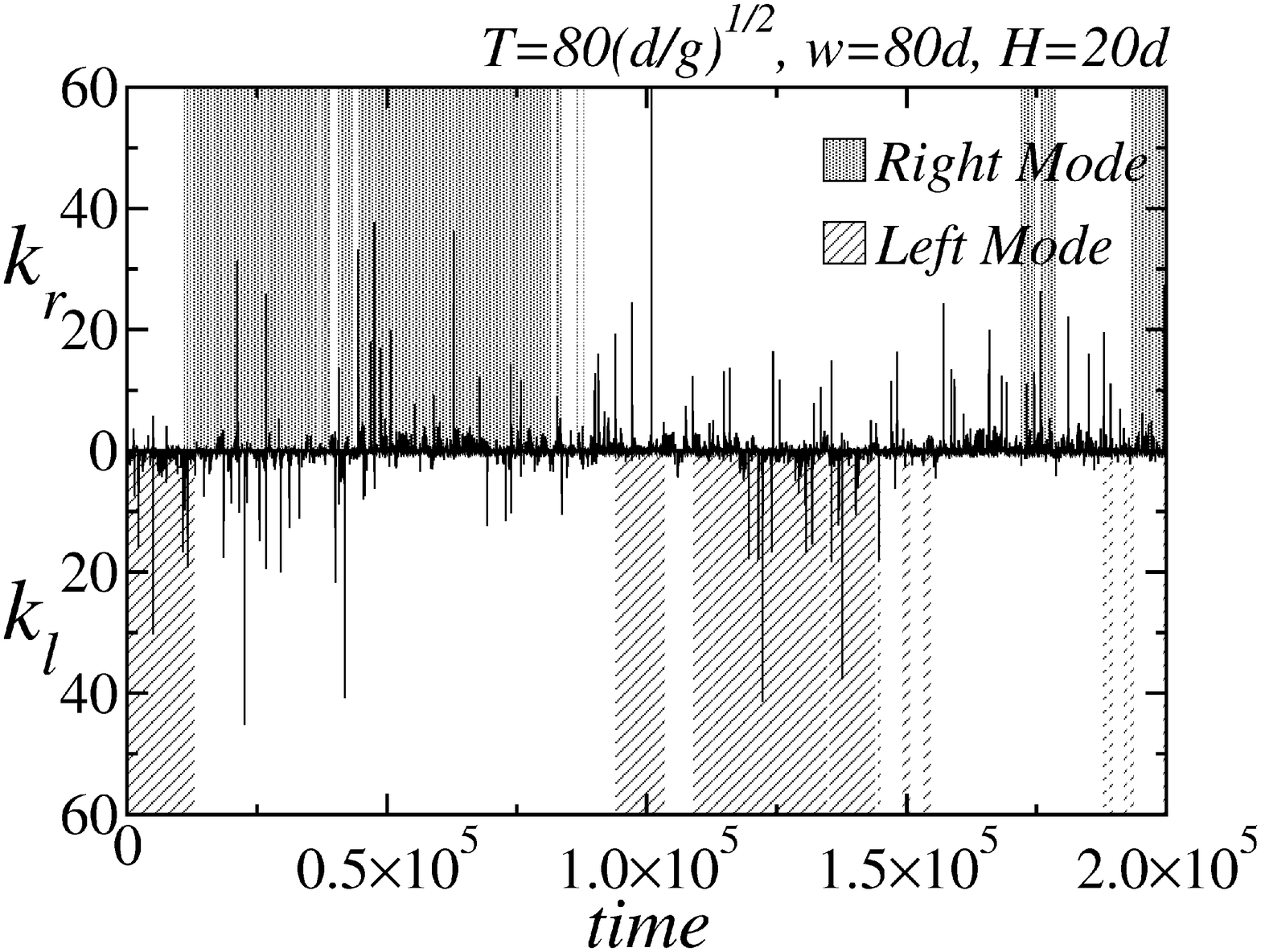}
\caption{\label{tklkr} The time series of $k_{l}$ and $k_{r}$ in the cases
  where $(w,H,T)=(80d,20d,2\sqrt{d/g})$ and $(80d,20d,80\sqrt{d/g})$ are
  plotted. The newly-defined mode is shown with $\Delta t=50T$.}
\end{center}
\end{figure}

%----------------------

We redefine the mode using $K'$ defined as follows.
\begin{eqnarray}
K'(t)=\left\{
\begin{array}{cc}
1, \quad & \mbox{if $k_{r}(t)>k_l(t)$,} \\
0, \quad & \mbox{if $k_{r}(t)=k_{l}(t)$,} \\
-1, \quad & \mbox{if $k_{r}(t)<k_{l}(t)$.}
\end{array}
\right.\label{newK}
\end{eqnarray}
The term, $K'=0$, is needed in the case where $T$ is large.
We define newly the right(left) mode as the state that the time for
$K'=1$($K'=-1$) amounts to more than $\gamma \Delta t$ in the time
$[t-\Delta t/2,t+\Delta t/2]$ where $0.5<\gamma<1$, and  $\Delta t$ is a
sufficiently large constant in comparison with $T$.
In addition, we introduce the competitive mode defined as the state that
the fractions of $K'=1$ and $K'=-1$ are comparable.

%----------------------

In Fig.\ref{tklkr}, we show the time series of $k_{l}$ and $k_{r}$ and
the newly-defined mode with $\gamma =0.8$ for $T=2\sqrt{d/g}$ and with
$\gamma =0.6$ for $T=80\sqrt{d/g}$.
Long-lived modes are observed not only for small $T$ but also for large
$T$.

%----------------------

\subsection{Continuation of the mode}

%----------------------

We infer that the mode continues because its memory is stored in either
of the shape of the sandpile or the motion of particles.
To determine which of them is more important factor, we carry out
examinations as follows.
We stop adding particles at a time and restart adding after waiting until
all particles cease, and we calculate the mode before stop and after
restart.
If the same mode tends to appear before stop and after restart, we
consider that its memory is stored in the shape because the motion of
particles ceases before adding particle is restarted.
Adding particles is restarted at the time when $k_{l}$ and $k_{r}$
decrease to a constant $k^{*}$.

%----------------------

This examination is repeated randomly, and the results show that the
memory is stored in the shape.
Table\ref{SM1} shows the results on the fraction of the mode after
restart in the case of the left(L) or right(R) mode before stop adding,
and Table\ref{SM2} shows that in the case of the competitive(C) mode, for
$(w,H,T)=(80d,20d,2\sqrt{d/g})$ where $\Delta t=50T$, $\gamma =0.8$, and
$k^{*}=1.0\times 10^{-6}mdg$.
Although the competitive mode appears more frequently than the left or right
mode in this criterion, the fraction that the same mode appears in
Table\ref{SM1} is higher than other modes.
In addition, also in Table\ref{SM2}, the fraction of the same mode is
significantly high.

%----------------------

\begin{table}
\begin{center}
\caption{\label{SM1}In the case of L or R before stop }
\begin{tabular}{cc} \hline
\quad After\textbackslash Before & L or R\\  \hline
Same & 58\%  \\
C & 40\%  \\
Contrary & 2\%  \\  \hline
total &  139 data \\ \hline
\end{tabular}
\caption{\label{SM2}In the case of C before stop}
\begin{tabular}{cc} \hline
\quad After\textbackslash Before & C  \\ \hline
 C &  72\% \\
 L or R &  28\% \\  \hline
 total & 211 data \\ \hline
\end{tabular}
\end{center}
\end{table}

%----------------------

\subsection{Relation between the fluctuation of the top location and the
  position of adding particles}

%----------------------

\begin{figure}
\begin{center}
\includegraphics[width=7cm]{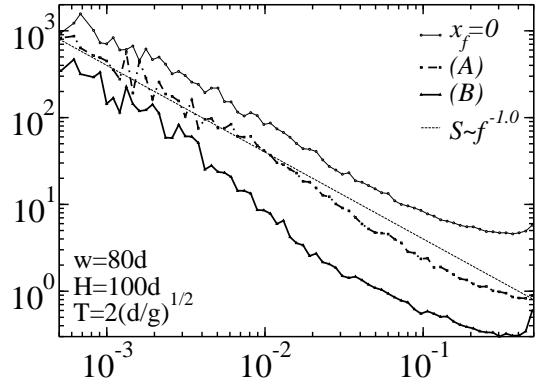}
\caption{\label{dwdl1} The power spectrum of $x_{top}$ for
 $(w,H,T)=(80d,100d,2\sqrt{d/g})$ in the cases (A), (B) and $x_{f}=0$. }
\end{center}
\end{figure}

\begin{figure}
\begin{center}
\includegraphics[width=7cm]{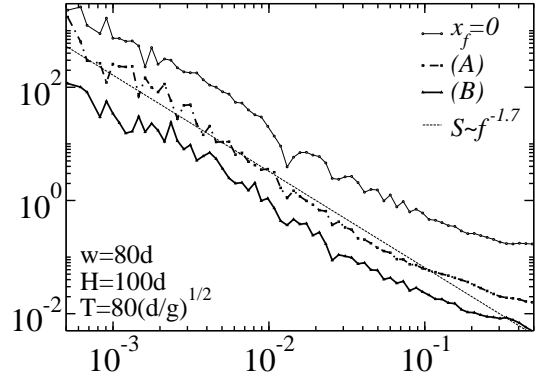}
\caption{\label{dwdl2}  The power spectrum of $x_{top}$ for
 $(w,H,T)=(80d,100d,80\sqrt{d/g})$ in the cases (A), (B) and $x_{f}=0$. }
\end{center}
\end{figure}

%----------------------

To clarify relations between the motion of the top and avalanches, we
change the position of adding particles because we infer that the
relations change with the position.
We choose randomly the horizontal position at which particles are
dropped, $x_{f}$, in a range and carry out experiments for the two cases
(A) and (B); (A) is the case where particles are fed uniformly above the
whole table, $-w/2<x_{f}<w/2$, and (B) is the case where we omit a
vicinity of the top from the rang, $w_{e}<|x_{f}|<w/2$ where $w_{e}$ is
a constant.
In both cases, we calculate the power spectrum of $x_{top}$ for
$(w,H,T)=(80d,100d,2\sqrt{d/g})$ and $(80d,100d,80\sqrt{d/g})$ and
compare with that in the case $x_{f}=0$.

%----------------------

In the cases of $x_{f}=0$ and (A), the exponents of the power spectra
are equal in a low frequency range.
The thick dashed lines in the Fig.\ref{dwdl1} and Fig. \ref{dwdl2}
show the power spectra in the case (A).
For reference, in the case $x_{f}=0$, we plot the power spectra (thin
solid lines) and the power functions with the exponent which is the same
with that of the power spectra (thin dashed lines), respectively.

%-----------------------

In addition, for small $T$, the exponent of the power spectrum depends
on whether particles are fed near the top or not.
The thick solid lines in Fig.\ref{dwdl1} and Fig. \ref{dwdl2} indicate
the power spectra in the case (B) where $w_{e}=10d$ because
$-10d<x_{top}<10d$ in the case $x_{f}=0$.
Although the exponents in the cases (B) and $x_{f}=0$ are almost equal
for $T=80\sqrt{d/g}$, the exponents in the case (B) is smaller than that
in the case $x_{f}=0$ for $T=2\sqrt{d/g}$.

%--------------------------

For small $T$, we consider that the motion of the top is different in the
cases (B) and $x_{f}=0$ because avalanches change with the position of
adding particle.
Avalanches are frequently accelerated by the impact of fed particles for
small $T$, and we infer that the probability
of the acceleration decreases with the distance between the landing position
of fed particles and the top because the landing position approaches
downstream of avalanches.
Therefore, the motion of the top changes with the distance.
Contrastingly, for large $T$,  in both cases $x_{f}=0$ and (B), the
probability is low because avalanches induced by a fed particle cease
before the next particle is fed, hence there is no difference in the
motion of the top.

%---------------------------

\section{ Fluid-Like State in the Surface of a Sandpile}

%---------------------------

\begin{figure}
\begin{center}
\includegraphics[width=7.5cm]{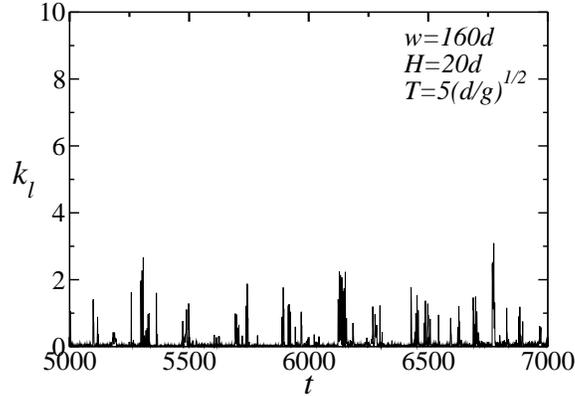}
\caption{\label{klw160} The time series of $k_{l}$ for $(w,H,T)=(160d,
 20d,5\sqrt{d/g})$}
\end{center}
\end{figure}

%---------------------------

\begin{figure}
\begin{center}
\includegraphics[width=7.5cm]{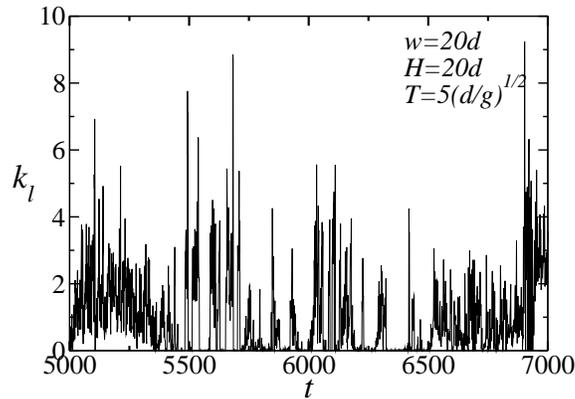}
\caption{\label{klw020} The time series of $k_{l}$ for
 $(w,H,T)=(20d,20d,5\sqrt{d/g})$}
\end{center}
\end{figure}

%---------------------------

In this section, we try to obtain a quantitative relation between the
states of the surface of a sandpile and the exponent of the power
spectrum of $x_{top}$, $\alpha$.
We infer that the state is characterized by some time scales for
avalanches, and that the motion of the top is related to the time
scales because the top is moved mainly by avalanches.

%----------------------------

The surface state and $\alpha$ depend on not only $T$ but also
$w$.
Actually, $\alpha$ for $(w,H,T)=(160d,20d,5\sqrt{d/g})$ is smaller than
that for smaller $w$ and the same $T$ as shown in Fig.\ref{a2D}.
As shown in Fig. \ref{klw160} and Fig. \ref{klw020}, $k_{l}$ for
$w=160d$ is clearly smaller than that for $w=20d$, and such small
kinetic energy is characteristic of the solid-like state, while the
state for $w=20d$ is fluid-like.

%----------------------------

\subsection{Time scales for avalanches}

%----------------------------

The surface state is related to two time scales for
avalanches.
One is the time required to cause an avalanche, and the other is
the lifetime of an avalanche.
In the case where the former is sufficiently larger than the latter,
the state is kept solid-like because the time between avalanches is
long.
Contrastingly, the state is fluid-like when these time scales are
comparable.

%----------------------

The lifetime of an avalanche, $T_{a}$, is independent of $w$.
Lifetime $T_{a}$ is calculated as the average of the duration
time in which $k_{l}$ or $k_{r}$ is kept larger than a constant $k_{a}$.
The duration time of an avalanche is well-defined when the feed rate is
small because each avalanche is plainly distinguishable.
Therefore, we calculate the time scales for large $T$.
Our numerical results with $k_{a}=0.05mdg$ show that $T_{a}$ is around
$5.0\sqrt{d/g}$ for $w=20d, 40d, 80d$ and $160d$ when
$(H,T)=(20d,80\sqrt{d/g})$.

%----------

\begin{figure}
\begin{center}
\includegraphics[width=7.5cm]{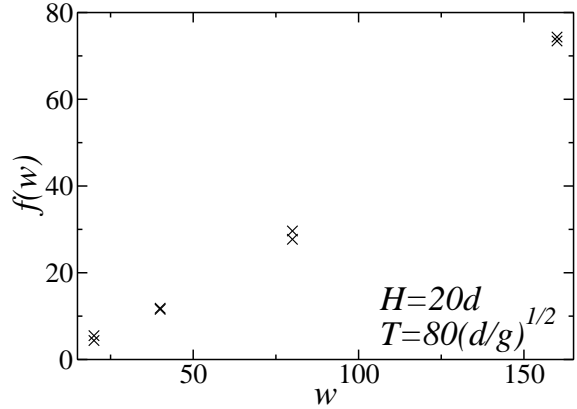}
\caption{\label{SD} We calculate $f(w)$ from the time series of $N_{l}$ or
 $N_{r}$ with the length $2.0\times 10^{6}\sqrt{d/g}$ for
 $(H,T)=(20d,80\sqrt{d/g})$.}
\end{center}
\end{figure}

The time required to cause an avalanche, $T_{s}$, depends on $T$ and $w$.
Time $T_{s}$ is defined as the time required to accumulate sufficient amount
of particles for causing an avalanche.
We postulate that $T_{s}$ is proportional to $T$, and $T_{s}$ is
represented as follows,
\begin{eqnarray}
T_{s}= Tf(w), \label{Tseq}
\end{eqnarray}
where $f(w)$ is the typical size of an avalanche and defined as the standard
deviation of $N_{l}(t)$ or $N_{r}(t)$, where $N_{l}(t)$ and $N_{r}(t)$ is
respectively the number of particles in the left half and right half of a
sandpile at time $t$.
The left half is defined as the part in the range $-w/2>x>-1.5d$, and
the right half is the part in the range $1.5d<x<w/2$.
We find that $f(w)$ increases with $w$ as shown in
Fig\ref{SD}.

%----------------------

The results show that the fluid-like state of the surface is kept for
a long time if $T$ and $w$ are small because $T_{a}$ and $T_{s}$ are
comparable, and that the state is solid-like if $T$ or $w$ is large
because $T_{a}\ll T_{s}$.

%----------------------

\begin{figure}
\begin{center}
\includegraphics[width=7.5cm]{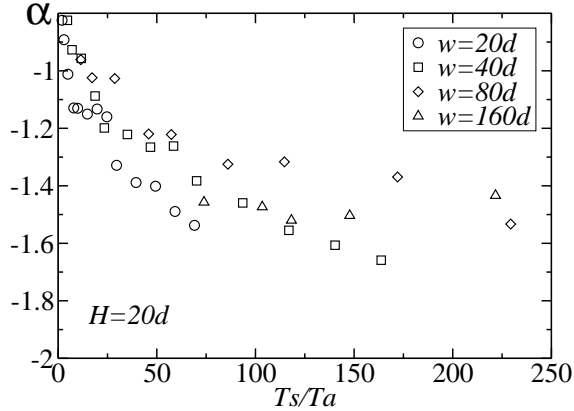}
\caption{\label{aTaTs} The dependence of $\alpha$ on
 $T_{s}/T_{a}$ when $H=20d$ and $T_{a}=5\sqrt{d/g}$}
\end{center}
\end{figure}

The exponent $\alpha$ is related to the surface state and a function of
$T_{s}/T_{a}$.
We assume that $\alpha$ depends on the ratio $T_{s}/T_{a}$ and rescale
the data in Fig.\ref{a2D} by $T^{*}=T_{a}/f(w)$.
The result is shown in Fig.\ref{aTaTs}.
However, to judge whether $\alpha$ depends on only $T/T^{*}$, more
elaborate simulation is needed to determine $\alpha$, $T_{a}$ and
$T_{s}$ more precisely.

%-----------------------

%-----------------------

%-----------------------

\section{Fluctuation of the Top Location and Avalanches in the
  Three-Dimensional Sandpile}

%-----------------------

\subsection{Dependence of fluctuation of the top location on $T$}

%----------------------

\begin{figure}
\begin{center}
\includegraphics[width=7.5cm]{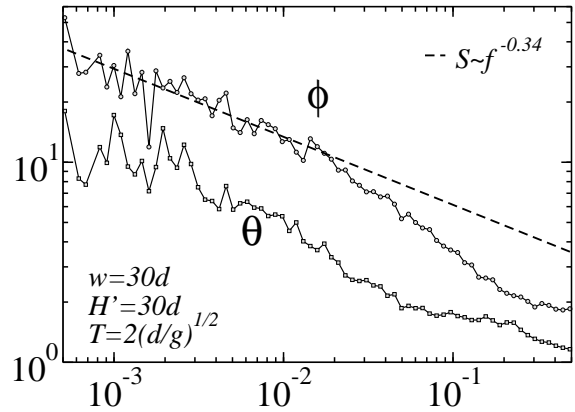}
\caption{\label{PS3D} Power spectra of $\phi$(upper solid line) and
  $\theta$(lower solid line) for $(w,H',T)=(30d,30d,2\sqrt{d/g})$. We move
  down the plot for $\theta$ in parallel to distinguish lines.}
\end{center}
\end{figure}

%----------------------

\begin{figure}
\begin{center}
\includegraphics[width=7.5cm]{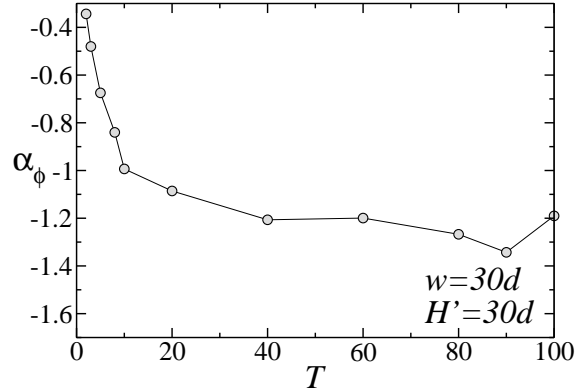}
\caption{\label{a3D} Dependence of $\alpha_{\phi}$ on $T$ for
  $(w,H')=(30d,30d)$.}
\end{center}
\end{figure}

%----------------------

In the three-dimensional sandpile, we measure the top location by the
cylindrical coordinates and calculate the power spectrum of the time series of
its azimuthal angle $\phi$ where $-\pi\leq\phi\leq\pi$ as in the same manner
for the two-dimensional sandpile.
The power spectrum obeys a power law, $S(f)\sim f^{\alpha_{\phi}}$, in a low
frequency range as shown in Fig.\ref{PS3D}.
Exponent $\alpha_{\phi}$ is obtained by a least-square fit of the double
logarithmic plot of the power spectrum in the frequency range $0.0005<f<0.01$.
Dependence of $\alpha_{\phi}$ on $T$ is shown in Fig.\ref{a3D}, which is
similar to Fig.\ref{a2D} for $\alpha$ in the two-dimensional sandpile,
although $\alpha_{\phi}$ is larger than $\alpha$.
We consider that a reason why $\alpha_{\phi} < \alpha$ is because $w$ is small
in the three-dimensional systems.

%----------------------

\subsection{Relation between the motion of the top and avalanches }

%------------------

We consider the direction of an avalanche projected on the horizontal
plain.
The direction is represented by the average of
particle momentum, $(\bar{p_{1}},\bar{p_{2}})$, which is defined by the
following equation,
\begin{eqnarray}
\bar{p_{l}}=\frac{1}{N}\sum_{i=1}^{N} m_{i}v_{i,l},\quad (l=1,2)
\end{eqnarray}
where $N$ denotes the number of particles on the table, and $v_{i,1}$ and
$v_{i,2}$ are $x$ and $y$ directional velocities of the $i$th particle,
respectively.
We define the direction of an avalanche as the azimuthal angle of the vector
$(\bar{p_{1}},\bar{p_{2}})$, $\theta$, where $-\pi \leq \theta \leq\pi$.

%------------------

To characterize the time series of $\theta$, we show its power spectrum in
Fig.\ref{PS3D}.
The power spectrum is proportional to that of $\phi$ in a low
frequency range.

%------------------

\section{Discussion}

%-----------------------

In our sandpile system and granular flow in a vertical pipe, there are
similar relations between the exponent of the power spectrum and the
phase space volume of each particle.
The exponent of the power spectrum of the top location depends on the
feed rate in the sandpile, and the power spectrum of the density wave in
the pipe obeys also a power law
\cite{
PengHerrmannPRE49(1994),
PengHerrmannPRE51(1995),
HorikawaNakaharaNakayamaMatsushitaJPSJ64(1995),
HorikawaIsodaNakayamaNakaharaMatsushitaPhysA233(1996),
NakaharaIsodaPRE55(1997),
MoriyamaKuroiwaMatsushitaHayakawaPRE80(1998),
HayakawaNakanishiPTPS130(1998),
MoriyamaKuroiwaIsodaAraiTatedaYamazakiMatsushitaTG01(2001),
YamazakiTatedaAwazuAraiMoriyamaMatsushitaJPSJ71(2002),
MoriyamaKuroiwaTatedaAraiAwazuYamazakiMatsushitaPrgTPS(2003),
HayakawaPRE72(2005)}
with the exponent which depends on the inflow rate to the pipe
\cite{NakaharaIsodaPRE55(1997),
YamazakiTatedaAwazuAraiMoriyamaMatsushitaJPSJ71(2002)}.
If the power spectrum $S_{d}$ obeys a power law, $S_{d}\sim f^{\beta}$,
the exponent $\beta$ increases with the volume in the phase space where
each particle able to move freely.
The phase space volume is decreased by restraint conditions which are
different in the sandpile and flow in the pipe.
In the sandpile, the volume in kinetic momentum space increases with the
feed rate.
In this case, avalanches occur frequently, and the surface state becomes
fluid-like.
In the pipe, the volume in kinetic momentum space and position space
are decreased with the inflow rate because clusters appear in the flow.
Developing these investigation, for granular systems, it is anticipated
that the relations between the local state, such as fluid-like or
solid-like, and the power spectrum in each systems are clarified
analytically.

%---------------------------------

\vspace*{1em}

%---------------------------------

We have investigated relations between the fluctuation of the top location
and avalanches in formation process of a sandpile using numerical simulations.
The top location is moved as particles are added one by one every
time interval $T$, and its power spectrum $S(f)$ obeys
a power law, $S(f)\sim f^{\alpha}$,  in a long time scale.
We found that the exponent $\alpha$ decreases with $T$ and the system size
$w$.

%---------------------------------

In a two-dimensional sandpile, we defined the right(left) mode as the state
that avalanches occur mainly on the right(left) slope of the sandpile,
and we found that the duration time of the left or right mode tends to
be long compared to $T$ because the memory of the mode is stored in the
shape of the sandpile.
In a three-dimensional sandpile, the direction of avalanches in the
horizontal plane changes with time, and the power
spectra of the top and the direction have the same exponent in a low
frequency range for small $T$.

%---------------------------------

The surface state of the sandpile and the exponent $\alpha$
depend on the ratio between the lifetime of an avalanche, $T_{a}$, and
the time required to cause an avalanche, $T_{s}$.
Our numerical results show that $T_{a}$ is a constant independent of
$w$, while $T_{s}$ increases with $w$ and $T$.
Therefore, the state is kept fluid-like when $T$ and $w$
are small because $T_{a}\sim T_{s}$, and it is solid-like when
$w$ or $T$ is large because $T_{a}\ll T_{s}$.
The state relates to the exponent $\alpha$, and we found
that $\alpha$ is a function of $T_{s}/T_{a}$.

\section*{Acknowledgment}

I appreciate helpful comments with Hisao Hayakawa, Hiroyuki Tomita,
Shinji Takesue, Mitsusada Sano and So Kitsunezaki.
The numerical calculations were carried out on Altix3700 BX2 at YITP in
Kyoto University.

\end{document}